\RequirePackage{amsmath}
\documentclass[runningheads]{llncs}
\usepackage[T1]{fontenc}
\usepackage{graphicx}
\usepackage{textcomp}
\usepackage{gensymb}
\usepackage{amssymb}
\usepackage{comment}
\usepackage[ruled, linesnumbered]{algorithm2e}
\usepackage{xcolor}
\usepackage{booktabs}
\usepackage{pdflscape}
\usepackage{xcolor}
\usepackage{caption}
\usepackage{subcaption}
\definecolor{webred}{rgb}{0.5,0,0}
\definecolor{webblue}{rgb}{0,0,0.8}
\usepackage[pdftex,colorlinks,citecolor=webblue,linkcolor=webred,]{hyperref}

\begin{document}
\title{Verification of Neural Network Control Systems in Continuous Time}
%
%
%\author{First Author\inst{1}\orcidID{0000-1111-2222-3333} \and
\author{Ali ArjomandBigdeli\inst{1} \and
Andrew Mata\inst{1} \and
Stanley Bak\inst{1}}
\institute{Stony Brook University\\ Stony Brook, NY 11794}
%\institute{Princeton University, Princeton NJ 08544, USA \and
%Springer Heidelberg, Tiergartenstr. 17, 69121 Heidelberg, Germany]

\maketitle              
\begin{abstract}
Neural network controllers are currently being proposed for use in many safety-critical tasks. 
Most analysis methods for neural network control systems assume a fixed control period.
In control theory, higher frequency usually improves performance.
However, for current analysis methods, increasing the frequency complicates verification.
In the limit, when actuation is performed continuously, no existing neural network control systems verification methods are able to analyze the system.
\vspace{1em}

In this work, we develop the first verification method for continuously-actuated neural network control systems.
We accomplish this by adding a level of abstraction to model the neural network controller. 
The abstraction is a piecewise linear model with added noise to account for local linearization error.
The soundness of the abstraction can be checked using open-loop neural network verification tools, although we demonstrate bottlenecks in existing tools when handling the required specifications.
We demonstrate the approach's efficacy by applying it to a vision-based autonomous airplane taxiing system and compare with a fixed frequency analysis baseline.
\end{abstract}
\section{Introduction}
Neural networks are increasingly gaining momentum for their potential use in safety-critical applications such as autonomous driving, medical treatment, and recommendation systems~\cite{Salman2020ANNinMed,pouyanfar2018suvey}. 
However, neural networks are complex, black-box function approximators that can have highly variable outputs when inputs are slightly perturbed.
Formal verification methods are commonly used to guarantee the absence of these adversarial perturbations that can cause significant performance degradation~\cite{goodfellow2015explaining,Eykholt2018RobustPhysical,gnanasambandam2021optical}. 

Most verification methods address the \emph{open-loop} neural network verification problem~\cite{brix2023vnncomp}.
In the open-loop problem, properties are expressed over inputs and outputs, and are proven for a given network.
Generally, the verification query can be reformulated as globally maximizing the value of the output given some bounded input constraints. 
However, due to the nonlinear activation function, open loop neural network verification can be difficult. 

\emph{Closed-loop} neural network verification, on the other hand, is often even more difficult. 
Closed-loop analysis aims to prove properties of a system consisting of a feed-back loop between a neural network controller and some system dynamics, usually expressed using differential equations~\cite{ARCH23}.
One strategy is time-bounded reachability, which unrolls the control loop and analyzes multiple executions of the neural network interacting with the physics.
As the control frequency increases, the number of steps needed to analyze the same amount of time goes up, creating a more difficult verification problem and, at some point, causing current verification techniques to fail.

Another verification method for closed-loop properties can reason over infinite execution by constructing a discrete abstraction of the system behavior.
Discrete abstractions allow reasoning over infinite-time closed-loop executions.
These methods may split the state space into a finite number of cells. 
Transitions between cells are defined using time-bounded reachability.
However, these methods also run into scalability issues.
Image-based controllers, for example, are high dimensional, adding further complexity to the verification problem. 
An additional challenge is mathematically formulating the set of possible images the system may encounter.
The most recent approach~\cite{katz2021verifiGAN} circumvents this issue by using a conditional generative adversarial network (cGAN) as a surrogate model for the real-world perception system. 
However, there are some important system characteristics that were not considered.
In their work, the control frequency runs at 1Hz. 
Using this approach, as the control frequency is increased, counter-intuitively, the area in the state space that is verified as safe is reduced~\cite{feiyang2024Scalable}.
The source of this is the abstraction error, which accumulates at every step in the analysis.
Another approach that verifies safety properties of the closed-loop behaviour for image-based controllers showed some interesting qualities of different frequency controllers. Notably, their results show that higher frequency controllers have smaller areas of guaranteed safety. In control theory, a higher frequency is usually associated with better control performance. This directly competes with verification: a lower frequency controller can be verified over more area of the state space. This motivates this work, which studies how to verify closed-loop behaviour with infinite frequency, the most accurate possible model. 
In this paper, we use the term infinite frequency and continuously actuated interchangeably.%

\begin{figure*}[t]
    \centering
    \includegraphics[width=1.0\textwidth]{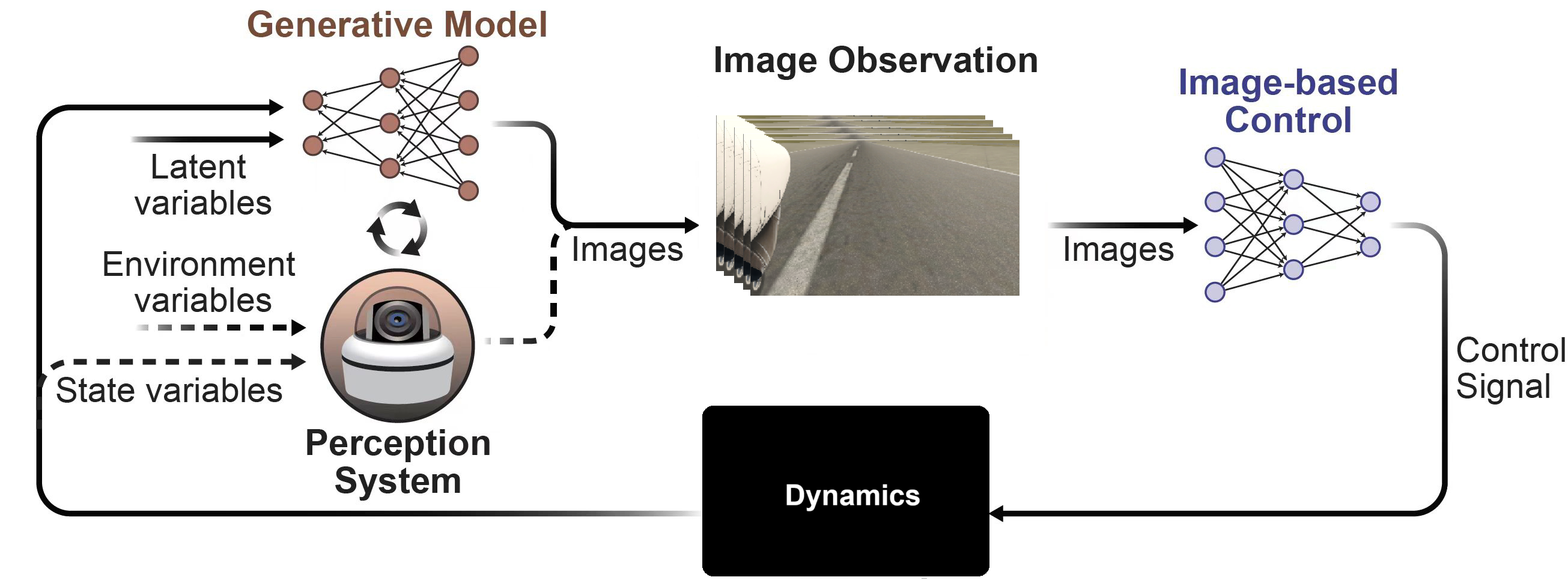}
    \caption{Autonomous Aircraft Taxiing System closed-loop diagram. In this work, we verify the performance of the infinite frequency version of this system, whereas prior work assumed a fixed frequency (1Hz).}
    \label{fig:taxinetcloseloop}
\end{figure*}

Infinite frequency introduces additional complexity for reachability analysis methods. 
We approach this issue by introducing another level of abstraction, approximating the neural network controller with a set of linear models that are valid in a local abstraction region.  
We provide safety guarantees by adding an additional uncertainty to each local model, ensuring the model is sound.
Reachability analysis can then be used for a timestep by taking advantage of the discrete abstraction of the state space to evaluate closed-loop safety properties.
We perform a case study on the Autonomous Aircraft Taxiing System (AATS)~\cite{katz2021verifiGAN,feiyang2024Scalable} and provide analysis on two closed-loop properties. The AATS can be seen in Figure~\ref{fig:taxinetcloseloop} and we provide more details in the following sections.
This is the first study, as far as the authors know, that considers continuously actuated neural network control systems for verification.

The main contributions of this work are as follows: 
\renewcommand{\labelitemi}{$\bullet$}
\begin{itemize}
    \item We demonstrate the current limitations of state-of-art verification tools on systems with high frequency controllers.
    \item We propose a method that generates local linear models combined with reachability analysis to analyze controllers with infinite frequency.
    \item We apply our framework on a complex vision neural network controller and find a promising verifiable region. Throughout the analysis, we also find challenging verification queries for state-of-art open-loop neural network verification tools. 
\end{itemize}

\section{Background}
This section provides an overview of the different methodologies and challenges involved in verifying open-loop neural network and neural network controllers, focusing on both fixed versus continuous actuation and state-based versus image-based verification approaches.

\subsection{Fixed vs. Continuous Actuation}
In control systems, particularly those involving neural networks, the assumption of a fixed frequency controller simplifies the analysis and design process. 
The prior work on our case study~\cite{katz2021verifiGAN}, for example, assumed a fixed frequency 1Hz controller.
Increasing the control frequency typically leads to better performance in control systems.
This is because a higher frequency allows the controller to update its outputs more frequently and act on newer sensor data. 
This results in more accurate and responsive control actions, as the system can quickly adjust to changes or disturbances. However, recent studies~\cite{feiyang2024Scalable} have shown that raising the frequency \emph{decreases} the area in state space where safety can be guaranteed. So, in practice, existing methods fail to verify neural network control systems running at higher frequencies.
This occurs because increasing the control frequency results in a higher number of steps required to analyze during the same time period. Consequently, the abstraction error accumulates over time, complicating the verification process.

\subsection{Open-loop Neural Network Verification}
Open-loop neural network verification~\cite{Kochdumper2023openclose} tries to verify that the output of a neural network, when applied to a specified set of inputs, meets a certain criteria. %
For a neural network represented as \( f_{NN}: \mathbb{R}^n \rightarrow \mathbb{R}^m \), with input set \( X \subseteq \mathbb{R}^{n} \) and unsafe output set \( P \subseteq \mathbb{R}^{m} \), open-loop neural network verification attempts to demonstrate that the output $f_{NN}(x)$ for all permissible inputs \( x \in X \) does not intersect with the unsafe set, ensuring that \( f_{NN}(x) \cap P = \emptyset \).

\subsection{State-based Neural Network Controller Verification} 
\label{sec:sbnnc}
This work aims to assess the performance of closed-loop neural network controllers in order to provide guarantees on a set of given safety properties. Several methods exist for verifying state-based neural network controller performance~\cite{Katz2021probControl,Julian2019NNACAS,huang2019reachnn,xiang2018reachableNNDynamicSys} and are further elaborated upon in Section~\ref{related_work}. These approaches use a combination of \emph{reachability analysis} and \emph{neural network verification} tools to reason about closed-loop safety properties. 
Reachability analysis, when applied to a system of differential equations, computes the set of states the system can enter after some bounded amount of time~\cite{schupp2019reachability}. In other words, the reachable set is the union of all possible states that a system can enter from a given set of initial states, subject to the system's dynamics and constraints~\cite{Althoff2021SetProp}.

We define a neural network controller $\mathcal{N_C}$ representing a policy $\pi$, as a function that maps states in a bounded state space $\mathcal{S}$ to an action in the action space $\mathcal{A}$. We consider a discrete-time dynamics model $s_{t+1} = f^\delta(s_t; a_t)$ that maps state-action pairs to the next state after $\delta$ time has passed.
By changing $\delta$ and keeping the control policy the same, we can analyze the performance of the controller under different fixed frequencies.
We can also consider the control in the limit as $\delta$ approaches zero, which is the infinite frequency case.

In this paper, we perform a discrete abstraction in which the state space $\mathcal{S}$ is partitioned in a finite number of hyperrectangular cells $c \in \mathcal{C}$ in order to make verification tractable. The set of feasible actions in a cell, denoted as $\mathcal{A}_c$, can be found with a neural network verification tool. We define an over-approximated dynamics model $f^{*}(c; \mathcal{A}_c)$ based on these specifications. The model maps each cell to a group of reachable next cells $\mathcal{C'}$ as well as their corresponding action ranges.

There are different types of reachability analysis: Backward reachability analysis starts from a target state and explores all possible paths leading to that state. Forward reachability analysis starts from an initial state and explores all reachable states through possible transitions.

% \subsection{Discrete Abstraction of Domain}

\subsection{Image-based Neural Network Controller Verification}

Real-world deployment of image-based neural network controllers remains difficult. There are no current, established methods for mathematically modeling the wide range of possible real-world images that a system might encounter. The high-dimensional nature of image input compounds this challenge in contrast with the typically low-dimensional states used in state-based neural network controllers. This makes not only verification but also thorough analysis of perception systems difficult. 

One current approach to these challenges is to substitute the real-world perception system with a generative model. This allows image-based neural network controllers to be formally verified~\cite{katz2021verifiGAN}. The state-of-the-art uses a conditional generative adversarial network (cGAN)~\cite{mirza2014cgan} that is trained to approximate the perception system and generates images based off low-dimensional system states alone. The concatenation of the cGAN and controller results in a unified neural network controller with low-dimensional state inputs. This enables the use of state-based closed-loop methods which are described in Section~\ref{sec:sbnnc}. Additionally, many safety properties are expressed using state variables rather than the image space.    

\section{Methodology}
The aim of this work is to perform verification on closed-loop neural network control systems whose controller runs at an infinite frequency. We use techniques, such as reachability analysis, that analyze a system's behavior in continuous-time. Reachability algorithms work with ODEs, neural network verification algorithms work on single inputs and outputs, so neither can directly be used for such analysis.
We navigate this issue by adding another layer of abstraction to our model. Namely, our method approximates the neural network controller with a piecewise linear model in the form of a set of linear models, each of which is valid for a local region. This allows traditional reachability tools to be used.

\begin{figure}[t]
    \centering
    \begin{subfigure}[b]{.42\textwidth}
        \centering
        \includegraphics[width=\textwidth]{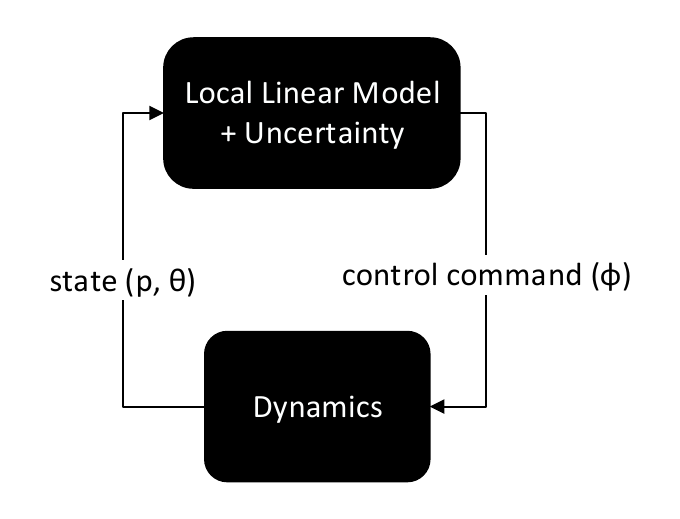}
        \caption{Nominal model.}
        \label{fig:nominal}
    \end{subfigure}
    % Leave a blank line between to stack them vertically
    \begin{subfigure}[b]{.85\textwidth}
        \centering
        \includegraphics[width=\textwidth]{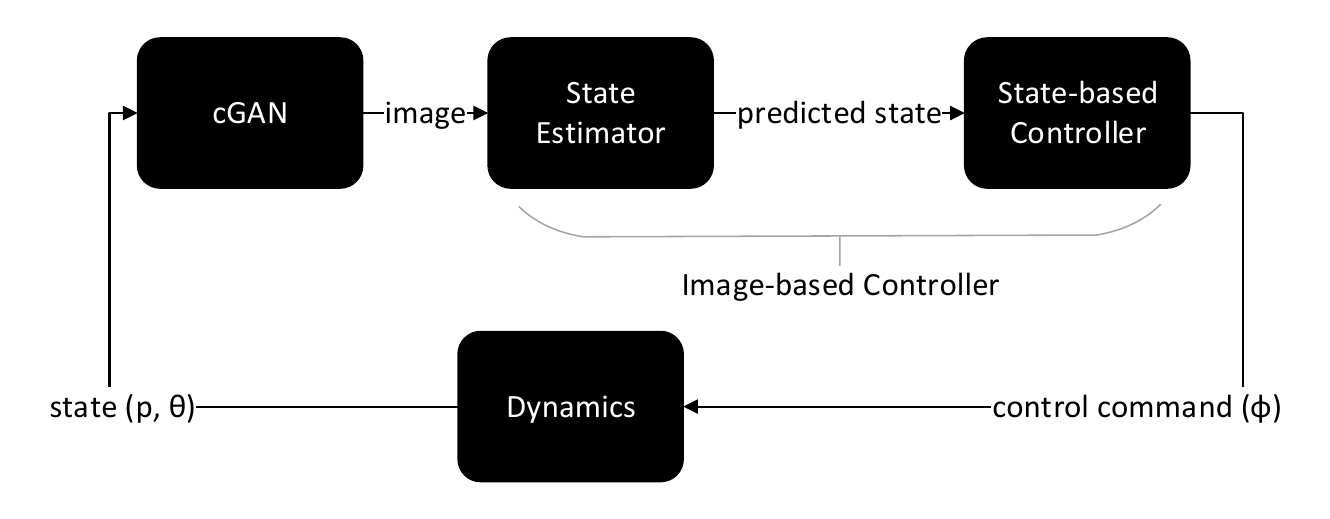}
        \caption{Real model.}
        \label{fig:real}
    \end{subfigure}
    \caption{The nominal model assumes an ideal, predictable system, whereas the real model includes disturbances from the environment. Conformant synthesis uses the nominal model and adds disturbances to tackle the real model's uncertainties.}
    \label{fig:nominalvsreal}
\end{figure}

In order to perform verification and provide safety guarantees, the local linear models must be \emph{conformant} with the output from the neural network controller. 
Our method achieves conformity by adding uncertainty to each local model. This results in a conservative abstract model, for each region, that over-approximates the network behavior. 
The \emph{nominal} model represents an idealized version of the system, where all parameters and dynamics are assumed to be known and predictable.
The \emph{real} model, on the other hand, incorporates uncertainties, disturbances, and possibly unknown dynamics. In conformant synthesis~\cite{stefan2023Reach}, the nominal model addresses the uncertainty of the real model by adding a disturbance to account for a range of possible scenarios. In our application, the cGAN and neural network controller are used to mimic the behavior of the real-life system.
Figure~\ref{fig:nominalvsreal} illustrates the nominal and real models. 

\begin{figure}[t]
  \begin{subfigure}{0.49\textwidth}
    \includegraphics[width=\linewidth]{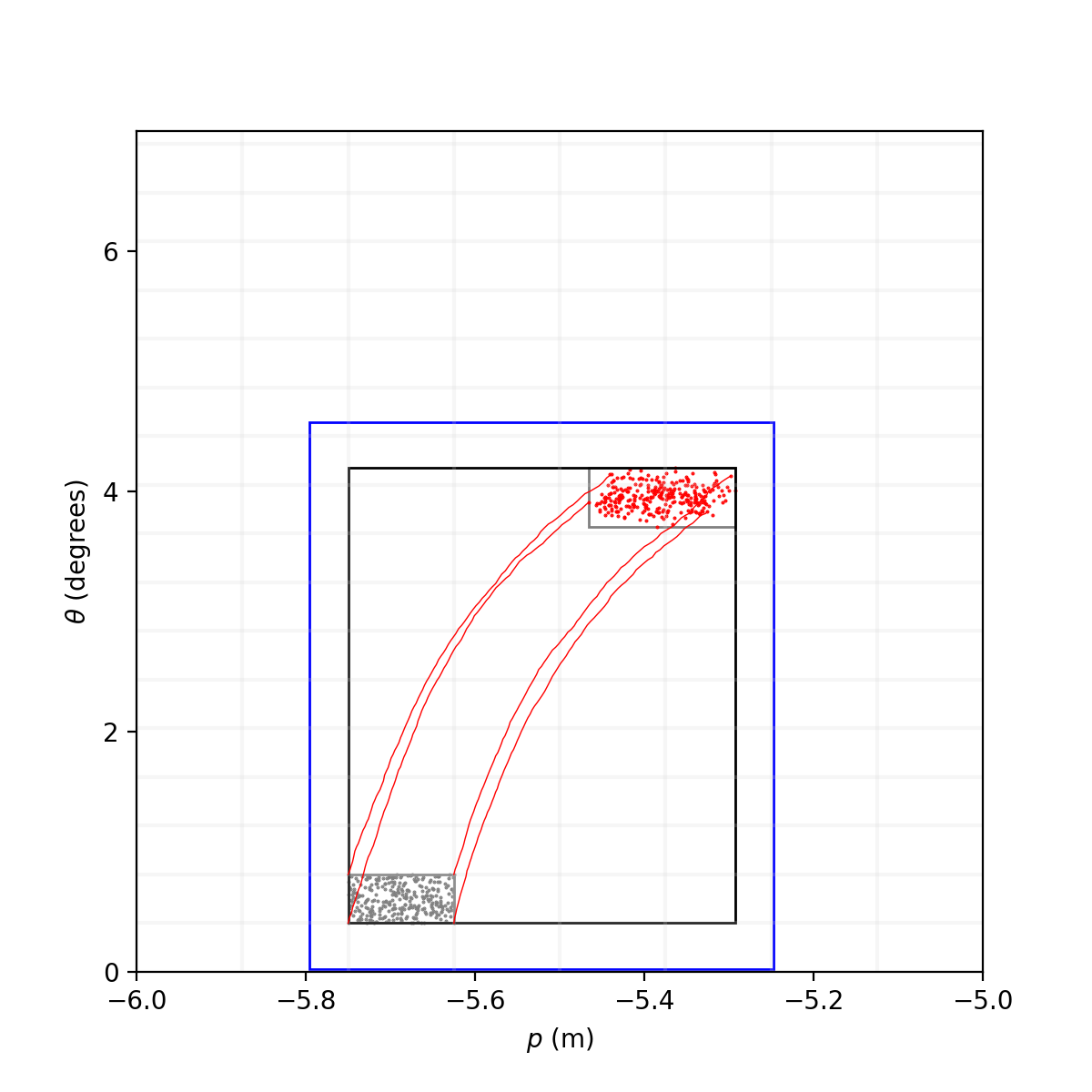}
    \caption{Abstraction region $\gamma$ in blue computed by simulation trajectories and bloating for a cell in the 2D state space of AATS. Gray dots represent sampled start states of simulations in the cell and the red dots are their final positions after a timestep.}
    \label{fig:onecell}
  \end{subfigure}%
  \hspace*{\fill} % maximize separation between the subfigures
  \begin{subfigure}{0.49\textwidth}
    \includegraphics[width=\linewidth]{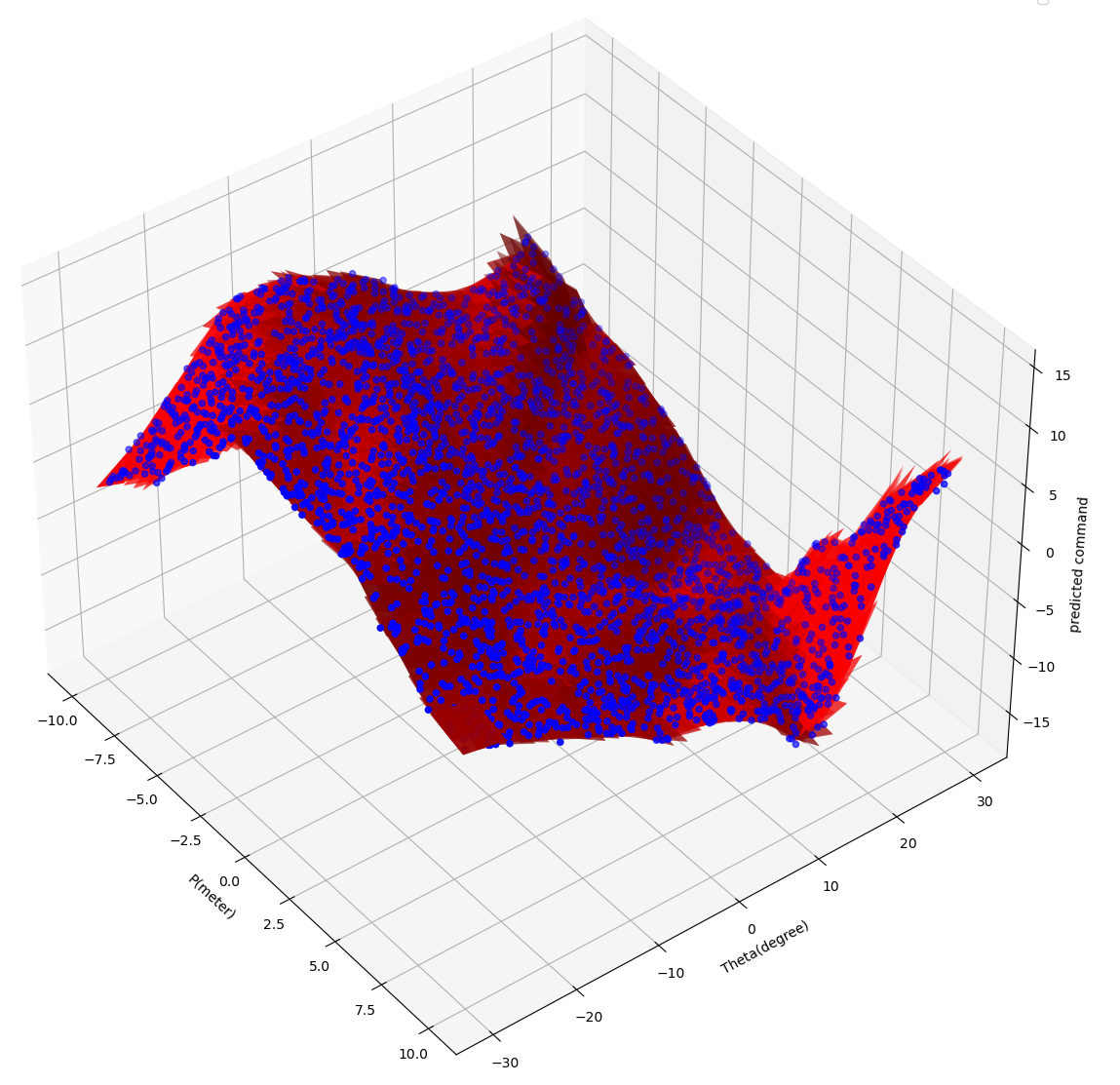}
    \caption{Control command prediction by state values for the entire state space. Set of linear models are in red and blue data points are neural network predictions.\\ \\}
    \label{fig:setoflinear}
  \end{subfigure}%
  \caption{Abstraction region for one cell and fitted linear models on all cells.}
  \label{fig:methodfig}
\end{figure}

For each local abstraction region the linear models should be valid or, in other words, contain all behaviors of the neural network controller for a time-step of the system. However, the behavior of the system may drift out of the local region throughout an execution of the time-step. This requires the linear model to be valid for a larger area than the initial cell region used in reachability analysis. Rather, it needs to cover the areas that the real reachable set might occupy. Unfortunately, this situation becomes the chicken or the egg problem: finding the real reachable set is tricky without running reachability analysis on the linear model first.        
To tackle this issue, we use random simulation trajectories as a witnesses and consider an \emph{abstraction region} $\gamma$ that encloses the start and end positions of the sampled points. 
Additionally, to take into the account the case where simulations might not find the worst-case scenarios, we bloat the region by an additional percentage that we define as the \emph{bloating factor} $bf$.
This process is detailed in Figure~\ref{fig:onecell} for a particular cell. After obtaining the abstraction region for the cell, linear regression is done to obtain a fitted linear model for each cell, shown in Figure~\ref{fig:setoflinear}.
For each cell, we consider the maximum and minimum difference of the neural network output and the local linear model's output as the upper and lower bound of the \emph{uncertainty} $\mathcal{U}$ that we use in our conformant synthesis.

Given the set of locally valid linear models, we use existing tools to perform reachability analysis with these models to obtain the reachable sets after a given time-step. We take advantage of the discrete state space abstraction to find the finite reachable cells. We define a cell abstraction graph $\langle C, E \rangle$ where $C$ is the set of cells in the discrete state abstraction and the edges $E$ are the state transitions between cells. This makes it possible to reason over infinite-time closed-loop executions.  

Simulation alone, however, may be insufficient to ensure that simulation trajectories do not leave the abstraction region in the middle of a time-step. We address this issue by checking the containment of the intermediate reachable sets and check that they are inside the abstraction region. If the containment check fails, we continually increase the bloating factor and rerun the simulation with a new, larger local region until the containment check passes. 

Our method is presented in Algorithm~\ref{algo:reachable_set}. We begin with iterating through each cell $c \in \mathcal{C}$ and compute the weight $A_{mat}$ and bias $b$ of the the linear model for each cell denoted by $Linearize$ (line 5). $Linearize$ additonally returns an uncertainty $\mathcal{U}$ and abstraction region $\gamma$. We next compute the reachable cells after elapsing a time-step, shown with the operator $Reach$ (line 6). The operator $Reach$ returns the final reachable cells $C_f$ and the intermediate reachable cells $C_i$ throughout the time-step. The intermediate cells are used for containment checking to ensure the soundness of the linear model in the abstraction region $\gamma$ (lines 8-12). After obtaining the cell abstraction graph $\langle C, E \rangle$,  we have the possible transitions from all cells. 

We then perform forward and backward reachablity on the abstraction graph to find the final reachable state. Algorithm~\ref{algo:forward_reachable_set} outlines the forward reachability process. We loop through and append the successor cells in the cell abstraction graph until the reachable cell set converges.     

For backward reachability, we compute the initial unsafe states by successor cells achieved by the cell abstraction graph that meets the unsafe condition. Then, we generate the predecessor abstraction graph by reversing the transitions in the cell abstraction graph. We pass the initial unsafe states and the predecessor abstraction graph into Algorithm~\ref{algo:forward_reachable_set} to find all cells that are unsafe.

\begin{algorithm}[t]
\caption{Compute Abstract Transitions (CAT)}
\label{algo:reachable_set}
\textbf{Inputs:} Network Controller $\mathcal{N_C}$, Cell Partition $\mathcal{C}$, Bloating Factor $bf$, Increasing Factor $inc$\;
\textbf{Outputs:} Cell Abstraction Graph $\langle C, E \rangle$\;
$E \leftarrow \emptyset$\;
\For{c in $\mathcal{C}$}{
    \tcp{generate linear model}
    $A_{mat}, b, \mathcal{U}, \gamma$ $\leftarrow$ $Linearize(\mathcal{N_C}, c, bf)$\;
    \tcp{get successor cells}
    $C_{f}$, $C_{i}$ $\leftarrow$  $Reach(A_{mat}, b, \mathcal{U}, \gamma)$\;
    \tcp{add new cell transitions to graph}
    $E \leftarrow E \cup \{(c, c_f) | c_f \in C_{f}\}$\;
    \tcp{check containment of intermediate cells}
    \While{$ \lnot ContainCheck(C_{i}, \gamma)$}{
        $bf$ $\leftarrow$ $bf$ * $inc$\;
        $A_{mat}, b, \mathcal{U}, \gamma$ $\leftarrow$ $Linearize(\mathcal{N_C}, c, bf)$\;
        $C_{f}$, $C_{i}$ $\leftarrow$  $Reach(A_{mat}, b, \mathcal{U}, \gamma)$\;
        $E \leftarrow E \cup \{(c, c_f) | c_f \in C_{f}\}$\;
    }
}
\end{algorithm}

\begin{algorithm}[t]
\caption{Forward Reachable Set Computation}
\label{algo:forward_reachable_set}
\textbf{Input:} Initial Cell Set $C_0$, Cell Abstraction Graph $\langle C, E\rangle$\;
\textbf{Output:} Reachable Cells $C_{end}$\;
$n \leftarrow 0$\;
\tcp{while reachable cells haven't converged}
\While{$C_{n+1} \neq C_{n}$}{
    $n \leftarrow n + 1$\;
    $C_{n+1}$ $\leftarrow$  $\emptyset$\;
    \For{c in $C_n$}{
        \tcp{get successor cells from abstraction graph}
        $C' \leftarrow \{c' | (c, c') \in E\}$\;
        \tcp{append successor cells}
        $C_{n+1}$ $\leftarrow$  $C_{n+1} \cup C'$\;
    }
}
\end{algorithm}

\section{Evaluation}
We evaluate our method on an autonomous aircraft taxiing system (AATS)~\cite{katz2021verifiGAN} that is based on Boeing’s TaxiNet~\cite{Staudinger2018Taxinet}. For the reachability analysis we used Continuous Reachability Analyzer (CORA)~\cite{Althoff2015CORA}. We provide a complete comparison between our proposed method and the existing baseline method~\cite{katz2021verifiGAN}.

\subsection{Autonomous Aircraft Taxiing System}
The goal of the Autonomous Aircraft Taxiing System (AATS) is to control an aircraft's taxiing at a steady speed on a taxiway. The inputs to the system are the images obtained from a camera positioned on the right wing of the aircraft~\cite{Julian2019NNACAS}. In this work, identically to the previous work~\cite{katz2021verifiGAN}, the goal is to verify that the aircraft does not violate the following safety properties: 
\begin{itemize}
    \item  Property 1 (P1): The aircraft stays within the runway and will not leave it. 
    \item  Property 2 (P2): The airplane will be directed toward the runway's center.
\end{itemize}

The aircraft's state is determined by its crosstrack position ($p$) and heading angle error ($\theta$) with respect to the taxiway center line. These state variables change in accordance with non-linear continuous-time dynamics, driven by the control signal $\phi$:

%%%% continuous-time dynamics:
\begin{equation}
\begin{split}
    \dot{p} = v \sin \theta \\
    \dot\theta = \frac{v}{L} \tan \phi
\end{split}
\end{equation}

\noindent where $L$ is the distance ($5 m$) between the front and back wheels, $v$ is the taxi speed ($5 m/s$)
%, $\Delta t$ is the time step
, and $\phi$ is the steering angle control input. An image-based controller generates the control signal $\phi$. A neural network predicts the state variables $p$ and $\theta$ and a proportional control method produces the control signal $\phi$ by the following law:

\begin{equation}
\phi = -0.74p - 0.44\theta
\end{equation}

\begin{figure*}[t]
    \centering
    \includegraphics[width=0.35\textwidth]{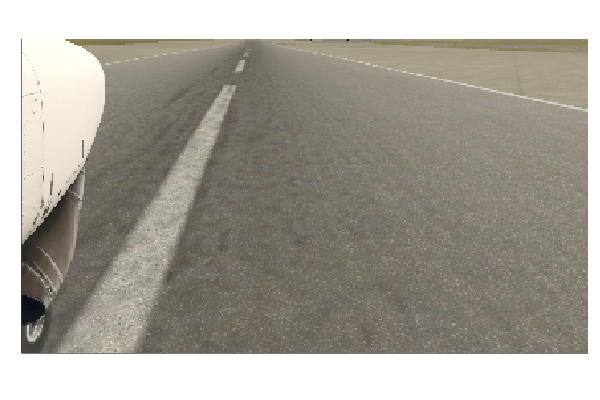}
    \includegraphics[width=0.35\textwidth]{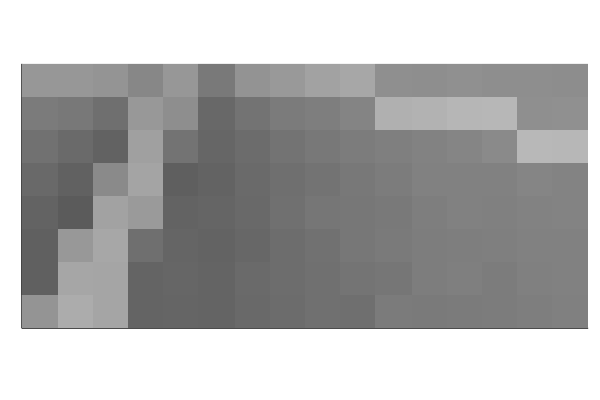}
    \caption{Example image of the runway taken from an aircraft wing-mounted camera (left) and its corresponding downsampled image (right). Images from~\cite{katz2021verifiGAN}.}
    \label{fig:taxiimg}
\end{figure*}

In previous work~\cite{katz2021verifiGAN}, the authors assumed a fixed frequency 1Hz controller. But in our work, we consider an infinite frequency. A cGAN is used to approximate the perception system, with the heading angle error ($\theta$) and the aircraft's crosstrack position ($p$) as inputs. To reflect fluctuations in the environment, two latent variables with a range of -0.8 to 0.8 are introduced. In the prior work, several simplifications were made to facilitate the verification of the surrogate system~\cite{katz2021verifiGAN}. Initially, the input RGB images, with dimensions of 200 × 360, were reduced to 8 × 16 grayscale images, which can be seen in Figure~\ref{fig:taxiimg}. This reduction allowed the implementation of a smaller feed-forward neural network controller with ReLU activations. Initially, a deep convolutional GAN (DCGAN) was used to train the cGAN. However, it was later replaced by a more compact feed-forward neural network that simulates the image generation capabilities of the DCGAN. The integrated network that encompasses both the cGAN and the image-based controller consists of 7 feed-forward layers, all equipped with ReLU activations. Figure~\ref{fig:taxinetcloseloop} shows AATS and its components.

\subsection{Closed-loop Verification}

\begin{figure}[t]
    \centering
    \begin{subfigure}[b]{1.0\textwidth}
        \centering
        \includegraphics[width=\textwidth]{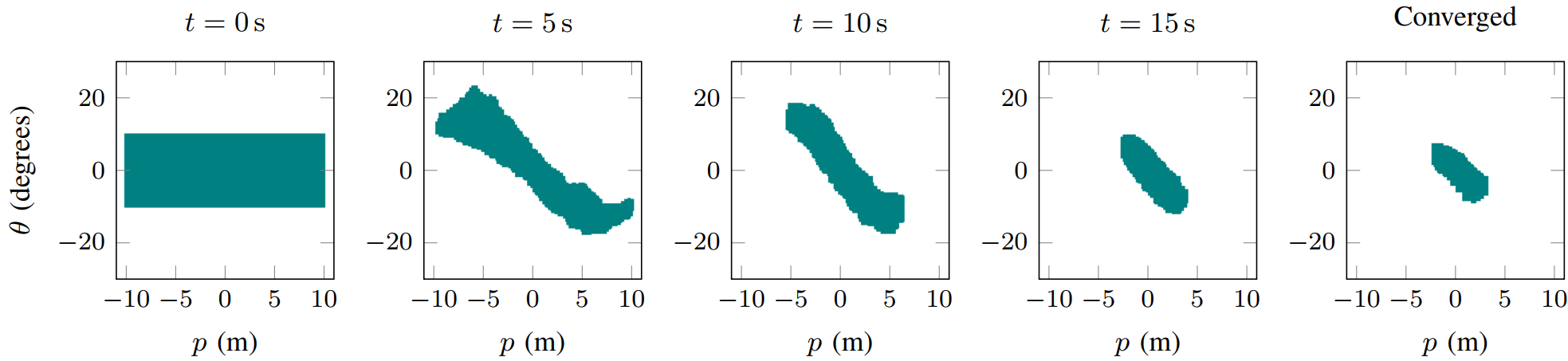}
        \caption{Baseline reachable set overtime for property 2 when starting from a state with $p \in [-10m, 10m]$ and $\theta \in [-10\degree, 10\degree]$. Converged at 20 seconds. Image from~\cite{katz2021verifiGAN}.}
        \label{fig:prop2baseline}
    \end{subfigure}
    % Leave a blank line between to stack them vertically
    \begin{subfigure}[b]{1.02\textwidth}
        \centering
        \includegraphics[width=\textwidth]{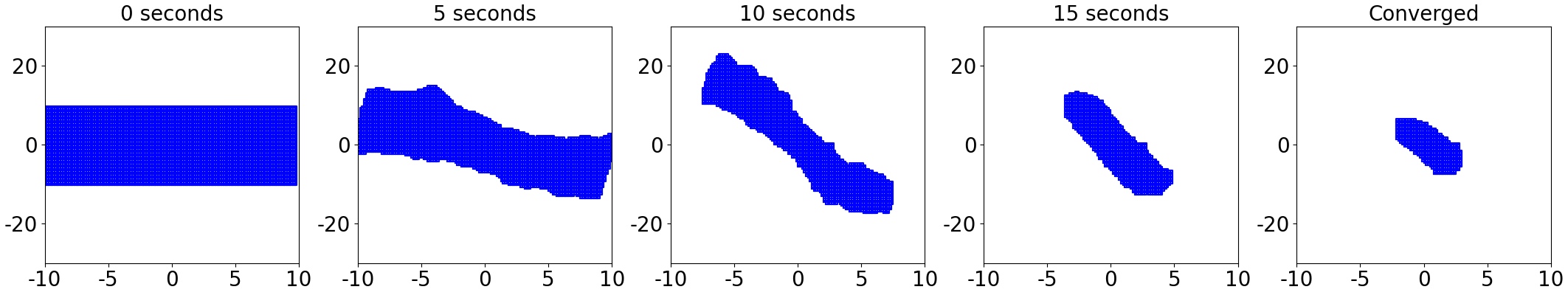}
        \caption{Reachable set overtime for property 2 generated by our modified approach at fixed 1Hz frequency setting when starting from a state with $p \in [-10m, 10m]$ and $\theta \in [-10\degree, 10\degree]$. Converged at 20 seconds.}
        \label{fig:prop2fixed1hz}
    \end{subfigure}
    \caption{Convergence of reachable sets over time. Comparison of the prior fixed frequency method and our modified approach at fixed 1Hz frequency setting using same initial states. Although the results are qualitatively similar, small differences are apparent due to the way physics are handled (the prior work uses a 20Hz time discretization), as well as how NN output bounds are obtained.}
    \label{fig:baselinevsoursprop2}
\end{figure}

\begin{figure}[t]
  \begin{subfigure}{0.2\textwidth}
    \includegraphics[width=\linewidth]{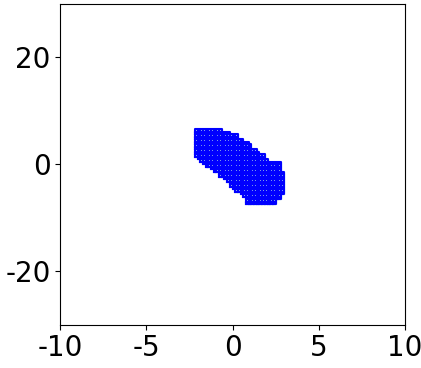}
    \caption{1Hz}
    \label{fig:prop2fixed1Hz_f}
  \end{subfigure}%
  \hspace*{\fill} % maximize separation between the subfigures
  \begin{subfigure}{0.2\textwidth}
    \includegraphics[width=\linewidth]{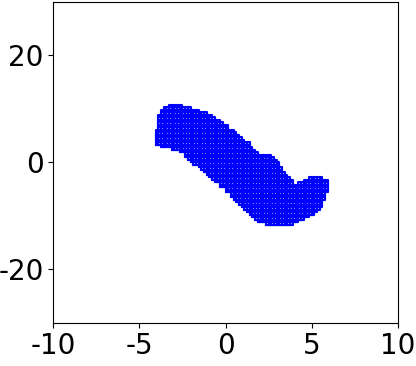}
    \caption{2Hz}
    \label{fig:prop2fixed2Hz}
  \end{subfigure}%
  \hspace*{\fill} % maximize separation between the subfigures
  \begin{subfigure}{0.2\textwidth}
    \includegraphics[width=\linewidth]{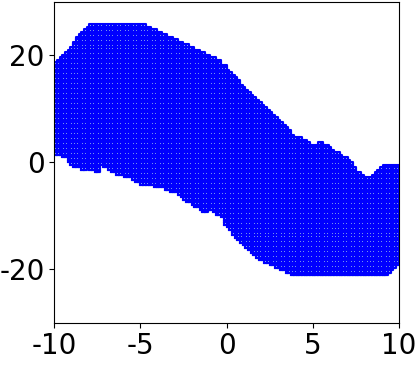}
    \caption{10Hz}
    \label{fig:prop2fixed10Hz}
  \end{subfigure}%
  \hspace*{\fill} % maximize separation between the subfigures
  \begin{subfigure}{0.2\textwidth}
    \includegraphics[width=\linewidth]{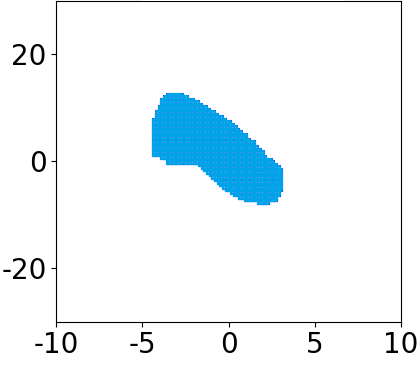}
    \caption{$\infty$Hz}
    \label{fig:prop2inf}
  \end{subfigure}%
  \hspace*{\fill} % maximize separation between the subfigures
  \begin{subfigure}{0.2\textwidth}
    \includegraphics[width=\linewidth]{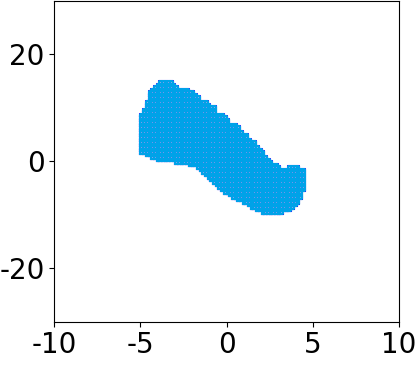}
    \caption{$\infty$Hz + $\mathcal{U}$}
    \label{fig:prop2infu}
  \end{subfigure}%
  \caption{Final reachable sets over time for property 2 in different settings. While increasing the fixed frequency leads to exploding, the infinite frequency using our method leads to convergence to a small area.}
  \label{fig:prop2all}
\end{figure}

After the computation of the cell abstract graph (in Algorithm~\ref{algo:reachable_set}), we apply the discrete abstraction and use both forward and backward reachability to perform closed-loop verification with respect to the safety properties. In the previous work~\cite{katz2021verifiGAN}, Katz at al. verify the safety properties by finding the upper and lower bounds of the neural network controller with a neural network verification tool. They also compute the upper and lower bound of the dynamics by solving an optimization problem, taking advantage of the monotonicity of AATS' dynamics. In order to use their method as a baseline, we modify our method to handle fixed frequency control, using $\mathcal{U}$ as the bounds of the neural network and run the reachability by passing the dynamics' ODE to the reachability tool. We ran our modified method with fixed frequency settings for several different frequencies (1Hz, 2Hz, 10Hz and 100Hz). Based on these results we draw conclusions about how well the fixed-frequency method works as the frequency is increased. 

As in the previous work, we define the discrete abstraction of the state space by dividing each dimension into 128 bins of identical size which in turn divides the space into 16384 uniform cells. For verifying Property 2, we compute forward reachability as described in Algorithm~\ref{algo:forward_reachable_set} and determine if the final reachable set converges. We first evaluate the Property 2 at 1Hz to replicate the fixed frequency of the baseline, which can be seen in Figure~\ref{fig:baselinevsoursprop2}. It also illustrates how the overapproximated reachable set changes over time. We see that the reachable set gets smaller until it converges into a small area in the middle of the state space. We know that the reachable set has converged if we find an invariant set through the forward reachability process. Next, we evaluate Property 2 with the initial set $p \in [-9 m, 9 m]$, $\theta \in [-10\degree, 10\degree]$ for different frequency including the infinite frequency. The final reachable sets can be seen in Figures~\ref{fig:prop2fixed1Hz_f}--\ref{fig:prop2fixed10Hz} for 1Hz, 2Hz and 10Hz fixed frequencies and in Figure~\ref{fig:prop2inf} for infinite frequency.  

\begin{figure}[t]
  \begin{subfigure}{0.25\textwidth}
    \includegraphics[width=\linewidth]{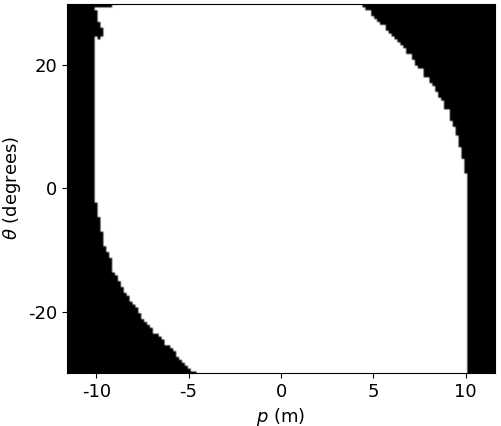}
    \caption{1Hz}
    \label{fig:prop1_1Hz}
  \end{subfigure}%
  \hspace*{\fill} % maximize separation between the subfigures
  \begin{subfigure}{0.25\textwidth}
    \includegraphics[width=\linewidth]{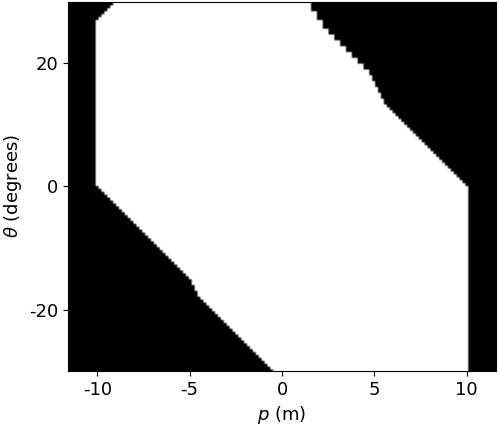}
    \caption{10Hz}
    \label{fig:prop1_10Hz}
  \end{subfigure}%
  \hspace*{\fill} % maximize separation between the subfigures
  \begin{subfigure}{0.25\textwidth}
    \includegraphics[width=\linewidth]{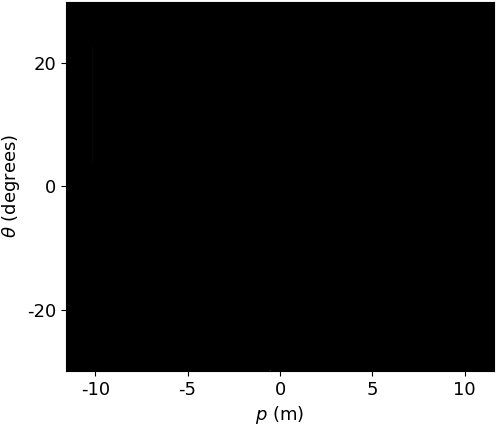}
    \caption{100Hz}
    \label{fig:prop1_100Hz}
  \end{subfigure}%
  \hspace*{\fill} % maximize separation between the subfigures
  \begin{subfigure}{0.25\textwidth}
    \includegraphics[width=\linewidth]{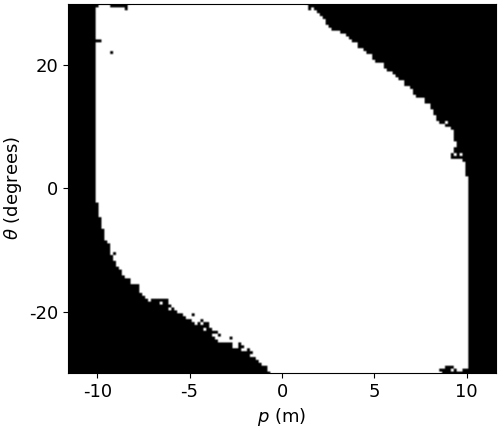}
    \caption{$\infty$Hz + $\mathcal{U}$}
    \label{fig:prop1withubloat}
  \end{subfigure}%
  \caption{Set of states guaranteed to meet the condition P1 with respect to the  space state. States colored white are assured to be safe, whereas states in black are undetermined. While increasing the fixed frequency lead to no verified area, with our method we find safe region for infinite frequency as can be seen in~\ref{fig:prop1withubloat}.}
  \label{fig:prop1all}
\end{figure}

For verifying Property 1, we compute backward reachability~\cite{Julian2019NNACAS} and explore all possible paths leading to unsafe states. Again like Property 2, we evaluate Property 1 for different fixed frequencies. Figures~\ref{fig:prop1_1Hz}--\ref{fig:prop1_100Hz} show the results for fixed frequencies of 1Hz, 10Hz and 100Hz; white cells indicate the controller will keep the airplane on the runway while black cells are unsafe regions where the aircraft may leave the runway. The image-based controller guarantees that the aircraft on the runway will maintain its position on the track if the plane starts from any crosstrack error within the runway limitations with a sufficiently small heading error. To have a better view, we also provide the quantitative results in Table~\ref{tab:prop1table}.

\begin{table}[t]
\centering
\begin{tabular}{ccccccc} % Three columns, all centered
\toprule % Top horizontal line
\textbf{Controller Frequency} & 1Hz & 2Hz & 10Hz & 100Hz & $\infty$Hz & $\infty$Hz + $\mathcal{U}$ \\ % Column names
\midrule % Middle horizontal line
\textbf{Verifiable \% Cells}    & 90 & 89 & 76 &  0 & 84 & 83    \\ 
\bottomrule % Bottom horizontal line
\end{tabular}
\vspace{5pt}
\caption{Percentage of verifiable area of the state space for AATS with different working frequency controllers. As frequency increases, the safe area decreases. Promisingly, our method finds a larger verifiable area than $\geq 10$Hz frequencies.}
\label{tab:prop1table}
\end{table}

\subsection{Verification of Abstraction Error}

We also check if the neural network output falls within the range we obtain from our linear model output plus some uncertainty. This is done using state-of-the-art neural network verification tools, including $\alpha$-$\beta$-CROWN~\cite{zhang2018efficient,xu2020automatic,xu2021fast,wang2021beta} and nnenum~\cite{bak2021nnenum}. 
The aforementioned tools, however, could not verify nor falsify the given verification queries for around 80\% of cells, either timing-out or throwing an exception. We plan to submit these queries as benchmarks to VNN-COMP~\cite{brix2023vnncomp}.

We hypothesize that the verification tools failed because the tools are generally used for verifying output properties, but, in these queries, our properties are over both the input and output space (given input $x \in X$ and output $Y = f_{NN}(x)$, the query is of the form: $A_{mat}\times X + b + \mathcal{U}_{lb} \leq Y \leq A_{mat}\times X + b + \mathcal{U}_{ub}$). 

Currently, tools do not accept input and output properties in a joint format. To use the tool $\alpha$-$\beta$-CROWN, we add a skip-connection layer to the neural network to have the input values to be present in the output space. For nnenum, we similarly had to create custom code to run the queries. We increased the timeout of the $\alpha$-$\beta$-CROWN to 12 hours but most of the cell queries timed out. A similar result occured with nnenum, with most cells throwing exceptions. 

For the cells that the verification queries succeeded on, most were verified and a few of them falsified by a counterexample that was around 5\% greater than the magnitude of our linear model plus some uncertainty. Thus, we decided to bloat further our uncertainty bounds by 10\% and rerun the experiments to try to verify the properties again. The final results were similar to those of our previous experiments, which can be seen in Figures~\ref{fig:prop2infu} for Property 2 and in the last column of Table~\ref{tab:prop1table} for Property 1.

\section{Related Work}
\label{related_work}
Verification of neural network control systems has been approached from several perspectives. 
One line of work combines neural network verification methods and reachability analysis. 
Many neural network verification methods and tools exist that use mixed integer linear programming (MILP) formulations ~\cite{ijcai2021p364,Fischetti2018DNNMIL,tjeng2019evaluating} or satisfiability module theory (SMT) encodings ~\cite{katz2017reluplex,katz2019marabou,guo2023occrob}. Marabou~\cite{katz2019marabou} is a parallel verifier that utilizes abstract interpretation and SMT solving.  
Other tools like nnenum ~\cite{bak2021nnenum} and NNV ~\cite{tran2020nnv} use set propagation and abstraction of reachable sets to achieve high verification performance. 
Bound Propagation-based algorithms ~\cite{zhang2018efficient,singh2019deeppoly,wang2018neurify,wong2018ConvexAdvPolytope} have also produced promising results.

Multiple methods exist for measuring the closed-loop performance of state-based neural network controllers~\cite{Katz2021probControl,Julian2019NNACAS,huang2019reachnn,xiang2018reachableNNDynamicSys}.
One strategy uses reachability analysis techniques along with the output of neural network verification tools. 
For closed-loop verification of image-based neural network controller, previous work~\cite{katz2021verifiGAN} used a generative model as a surrogate model to replace the real-world perception system. 
This reduces the closed-loop verification of image-based controllers to the verification of state-based controllers. 
Others~\cite{feiyang2024Scalable} reduced over-approximation errors by composition and unrolling. 
In one of their case studies, they also showed that higher frequency controllers have smaller verifiable areas of the state space.

Another line of work incorporates safety properties into the controller itself. 
One strategy modifies the training process of neural network controllers~\cite{platzer1,platzer2} to include safety when learning the environment.
Other works use safety shields if a controller's action will cause the system to reach an unsafe state~\cite{shield1,shield2}.

\section{Conclusion}
In this work, we present a method to perform verification on closed-loop neural network controllers in continuous time. The previous assumption of fixed frequency control systems is shown to be favorable for simplicity and stability analysis. However, uncertainties present in the real-world system introduce complexities that challenge this assumption. Furthermore, we demonstrated that current verification methods are ill-equipped to handle higher frequency control. Our method solves this issue and finds a reasonable area of guaranteed safety for continuous actuation compared to fixed-frequency methods.    

We perform a case-study on the image-based neural network control system AATS in which we verify two closed-loop safety properties. Although we demonstrate our method on an image-based network, it can be similarly applied to any state-based neural network control system that usually have simpler network architecture.

In practice, systems run at neither fixed nor continuous frequency. While most systems run at some fixed, discrete frequency, real-world errors can cause slight fluctuations to the working frequency. Thus, future work is necessary to investigate the impact of variable frequency control. We see this work as a step towards verification of multi-frequency controllers. This will further enhance the robustness of neural network controllers in unpredictable, real-world environments.       

\section{Acknowledgement}
This material is based upon work supported by the Air Force Office of Scientific Research and the Office of Naval Research under award numbers FA9550-19-1-0288, FA9550-21-1-0121, FA9550-23-1-0066 and N00014-22-1-2156, and the National Science Foundation under Award No. 2237229. Any opinions, findings, and conclusions or recommendations expressed in this material are those of the author(s) and do not necessarily reflect the views of the United States Air Force or the United States Navy.  

\bibliographystyle{splncs04}
\bibliography{bib}

\begin{thebibliography}{10}
\providecommand{\url}[1]{\texttt{#1}}
\providecommand{\urlprefix}{URL }
\providecommand{\doi}[1]{https://doi.org/#1}

\bibitem{Salman2020ANNinMed}
Al-Salman, O., Mustafina, J., Shahoodh, G.: A systematic review of artificial neural networks in medical science and applications. In: 2020 13th International Conference on Developments in eSystems Engineering (DeSE). pp. 279--282 (2020). \doi{10.1109/DeSE51703.2020.9450245}

\bibitem{shield1}
Alshiekh, M., Bloem, R., Ehlers, R., K{\"o}nighofer, B., Niekum, S., Topcu, U.: Safe reinforcement learning via shielding. In: Proceedings of the AAAI conference on artificial intelligence. vol.~32 (2018)

\bibitem{Althoff2015CORA}
Althoff, M.: An introduction to cora 2015. In: Proc. of the 1st and 2nd Workshop on Applied Verification for Continuous and Hybrid Systems. pp. 120--151. EasyChair (December 2015). \doi{10.29007/zbkv}, \url{https://easychair.org/publications/paper/xMm}

\bibitem{Althoff2021SetProp}
Althoff, M., Frehse, G., Girard, A.: Set propagation techniques for reachability analysis. Annual Review of Control, Robotics, and Autonomous Systems  \textbf{4}(Volume 4, 2021),  369--395 (2021). \doi{https://doi.org/10.1146/annurev-control-071420-081941}, \url{https://www.annualreviews.org/content/journals/10.1146/annurev-control-071420-081941}

\bibitem{bak2021nnenum}
Bak, S.: nnenum: Verification of relu neural networks with optimized abstraction refinement. In: Dutle, A., Moscato, M.M., Titolo, L., Mu{\~{n}}oz, C.A., Perez, I. (eds.) NASA Formal Methods. pp. 19--36. Springer International Publishing, Cham (2021)

\bibitem{brix2023vnncomp}
Brix, C., M{\"u}ller, M.N., Bak, S., Johnson, T.T., Liu, C.: First three years of the international verification of neural networks competition (vnn-comp). International Journal on Software Tools for Technology Transfer  \textbf{25}(3),  329--339 (2023)

\bibitem{Eykholt2018RobustPhysical}
Eykholt, K., Evtimov, I., Fernandes, E., Li, B., Rahmati, A., Xiao, C., Prakash, A., Kohno, T., Song, D.: Robust physical-world attacks on deep learning visual classification. In: 2018 IEEE/CVF Conference on Computer Vision and Pattern Recognition. pp. 1625--1634 (2018). \doi{10.1109/CVPR.2018.00175}

\bibitem{feiyang2024Scalable}
Feiyang, C., Bak, S.: Scalable surrogate verification of image-based neural network control systems using composition and unrolling (under review) (2024)

\bibitem{Fischetti2018DNNMIL}
Fischetti, M., Jo, J.: Deep neural networks and mixed integer linear optimization. Constraints  \textbf{23}(3),  296–309 (Apr 2018). \doi{10.1007/s10601-018-9285-6}, \url{http://dx.doi.org/10.1007/s10601-018-9285-6}

\bibitem{ARCH23}
Frehse, G., Althoff, M. (eds.): Proceedings of 10th International Workshop on Applied Verification of Continuous and Hybrid Systems (ARCH23), EPiC Series in Computing, vol.~96. EasyChair (2023)

\bibitem{platzer1}
Fulton, N., Platzer, A.: Safe reinforcement learning via formal methods: Toward safe control through proof and learning. In: Proceedings of the AAAI Conference on Artificial Intelligence. vol.~32 (2018)

\bibitem{platzer2}
Fulton, N., Platzer, A.: Verifiably safe off-model reinforcement learning. In: International Conference on Tools and Algorithms for the Construction and Analysis of Systems. pp. 413--430. Springer (2019)

\bibitem{gnanasambandam2021optical}
Gnanasambandam, A., Sherman, A.M., Chan, S.H.: Optical adversarial attack. In: Proceedings of the IEEE/CVF International Conference on Computer Vision. pp. 92--101 (2021)

\bibitem{goodfellow2015explaining}
Goodfellow, I.J., Shlens, J., Szegedy, C.: Explaining and harnessing adversarial examples. arXiv preprint arXiv:1412.6572  (2014)

\bibitem{guo2023occrob}
Guo, X., Zhou, Z., Zhang, Y., Katz, G., Zhang, M.: Occrob: efficient smt-based occlusion robustness verification of deep neural networks. In: International Conference on Tools and Algorithms for the Construction and Analysis of Systems. pp. 208--226. Springer (2023)

\bibitem{huang2019reachnn}
Huang, C., Fan, J., Li, W., Chen, X., Zhu, Q.: Reachnn: Reachability analysis of neural-network controlled systems. ACM Transactions on Embedded Computing Systems (TECS)  \textbf{18}(5s),  1--22 (2019)

\bibitem{Julian2019NNACAS}
Julian, K.D., Kochenderfer, M.J.: Guaranteeing safety for neural network-based aircraft collision avoidance systems. In: 2019 IEEE/AIAA 38th Digital Avionics Systems Conference (DASC). IEEE (Sep 2019). \doi{10.1109/dasc43569.2019.9081748}, \url{http://dx.doi.org/10.1109/DASC43569.2019.9081748}

\bibitem{katz2017reluplex}
Katz, G., Barrett, C., Dill, D.L., Julian, K., Kochenderfer, M.J.: Reluplex: An efficient smt solver for verifying deep neural networks. In: Computer Aided Verification: 29th International Conference, CAV 2017, Heidelberg, Germany, July 24-28, 2017, Proceedings, Part I 30. pp. 97--117. Springer (2017)

\bibitem{katz2019marabou}
Katz, G., Huang, D.A., Ibeling, D., Julian, K., Lazarus, C., Lim, R., Shah, P., Thakoor, S., Wu, H., Zelji{\'{c}}, A., Dill, D.L., Kochenderfer, M.J., Barrett, C.: The marabou framework for verification and analysis of deep neural networks. In: Dillig, I., Tasiran, S. (eds.) Computer Aided Verification. pp. 443--452. Springer International Publishing, Cham (2019)

\bibitem{katz2021verifiGAN}
Katz, S.M., Corso, A.L., Strong, C.A., Kochenderfer, M.J.: Verification of image-based neural network controllers using generative models. Journal of Aerospace Information Systems  \textbf{19}(9),  574--584 (2022)

\bibitem{Katz2021probControl}
Katz, S.M., Julian, K.D., Strong, C.A., Kochenderfer, M.J.: Generating probabilistic safety guarantees for neural network controllers. Machine Learning  \textbf{112}(8),  2903–2931 (Oct 2021). \doi{10.1007/s10994-021-06065-9}, \url{http://dx.doi.org/10.1007/s10994-021-06065-9}

\bibitem{Kochdumper2023openclose}
Kochdumper, N., Schilling, C., Althoff, M., Bak, S.: Open- and closed-loop neural network verification using polynomial zonotopes. In: Rozier, K.Y., Chaudhuri, S. (eds.) NASA Formal Methods. pp. 16--36. Springer Nature Switzerland, Cham (2023)

\bibitem{ijcai2021p364}
Kouvaros, P., Lomuscio, A.: Towards scalable complete verification of relu neural networks via dependency-based branching. In: Zhou, Z.H. (ed.) Proceedings of the Thirtieth International Joint Conference on Artificial Intelligence, {IJCAI-21}. pp. 2643--2650. International Joint Conferences on Artificial Intelligence Organization (8 2021). \doi{10.24963/ijcai.2021/364}, \url{https://doi.org/10.24963/ijcai.2021/364}, main Track

\bibitem{stefan2023Reach}
Liu, S.B., Sch{\"u}rmann, B., Althoff, M.: Reachability-based identification, analysis, and control synthesis of robot systems. arXiv e-prints pp. arXiv--2103 (2021)

\bibitem{mirza2014cgan}
Mirza, M., Osindero, S.: Conditional generative adversarial nets. arXiv preprint arXiv:1411.1784  (2014)

\bibitem{pouyanfar2018suvey}
Pouyanfar, S., Sadiq, S., Yan, Y., Tian, H., Tao, Y., Reyes, M.P., Shyu, M.L., Chen, S.C., Iyengar, S.S.: A survey on deep learning: Algorithms, techniques, and applications. ACM Comput. Surv.  \textbf{51}(5) (sep 2018). \doi{10.1145/3234150}, \url{https://doi.org/10.1145/3234150}

\bibitem{schupp2019reachability}
Schupp, S.: State set representations and their usage in the reachability analysis of hybrid systems. Ph.D. thesis, Dissertation, RWTH Aachen University, 2019 (2019)

\bibitem{singh2019deeppoly}
Singh, G., Gehr, T., Püschel, M., Vechev, M.: An abstract domain for certifying neural networks. Proceedings of the ACM on Programming Languages  \textbf{3}(POPL),  1–30 (Jan 2019). \doi{10.1145/3290354}, \url{http://dx.doi.org/10.1145/3290354}

\bibitem{Staudinger2018Taxinet}
Staudinger, T.C., Jorgensen, Z.D., Margineantu, D.D.: X-taxinet - an environment for learning and decision systems for airplane operations (2018)

\bibitem{tjeng2019evaluating}
Tjeng, V., Xiao, K., Tedrake, R.: Evaluating robustness of neural networks with mixed integer programming. arXiv preprint arXiv:1711.07356  (2017)

\bibitem{tran2020nnv}
Tran, H.D., Yang, X., Manzanas~Lopez, D., Musau, P., Nguyen, L.V., Xiang, W., Bak, S., Johnson, T.T.: Nnv: the neural network verification tool for deep neural networks and learning-enabled cyber-physical systems. In: International Conference on Computer Aided Verification. pp. 3--17. Springer (2020)

\bibitem{wang2018neurify}
Wang, S., Pei, K., Whitehouse, J., Yang, J., Jana, S.: Efficient formal safety analysis of neural networks. Advances in neural information processing systems  \textbf{31} (2018)

\bibitem{wang2021beta}
Wang, S., Zhang, H., Xu, K., Lin, X., Jana, S., Hsieh, C.J., Kolter, J.Z.: {Beta-CROWN}: Efficient bound propagation with per-neuron split constraints for complete and incomplete neural network verification. Advances in Neural Information Processing Systems  \textbf{34} (2021)

\bibitem{wong2018ConvexAdvPolytope}
Wong, E., Kolter, Z.: Provable defenses against adversarial examples via the convex outer adversarial polytope. In: International conference on machine learning. pp. 5286--5295. PMLR (2018)

\bibitem{xiang2018reachableNNDynamicSys}
Xiang, W., Lopez, D.M., Musau, P., Johnson, T.T.: Reachable set estimation and verification for neural network models of nonlinear dynamic systems. Safe, autonomous and intelligent vehicles pp. 123--144 (2019)

\bibitem{shield2}
Xiong, Z., Jagannathan, S.: Scalable synthesis of verified controllers in deep reinforcement learning. arXiv preprint arXiv:2104.10219  (2021)

\bibitem{xu2020automatic}
Xu, K., Shi, Z., Zhang, H., Wang, Y., Chang, K.W., Huang, M., Kailkhura, B., Lin, X., Hsieh, C.J.: Automatic perturbation analysis for scalable certified robustness and beyond. Advances in Neural Information Processing Systems  \textbf{33},  1129--1141 (2020)

\bibitem{xu2021fast}
Xu, K., Zhang, H., Wang, S., Wang, Y., Jana, S., Lin, X., Hsieh, C.J.: {Fast and Complete}: Enabling complete neural network verification with rapid and massively parallel incomplete verifiers. In: International Conference on Learning Representations (2021), \url{https://openreview.net/forum?id=nVZtXBI6LNn}

\bibitem{zhang2018efficient}
Zhang, H., Weng, T.W., Chen, P.Y., Hsieh, C.J., Daniel, L.: Efficient neural network robustness certification with general activation functions. Advances in Neural Information Processing Systems  \textbf{31},  4939--4948 (2018), \url{https://arxiv.org/pdf/1811.00866.pdf}

\end{thebibliography}
\end{document}